\documentstyle[aps,prb,preprint,epsf]{revtex}

\begin{document}

\bibliographystyle{aip}

\title{Magnetic susceptibility of EuTe/PbTe Heisenberg
antiferromagnetic superlattices: experimental and theoretical studies}

\author{Lorenzo Bergomi\footnote[4]{Author to whom
correspondence should be addressed, 
please use French address. \\
email: bergomi@spht.saclay.cea.fr}}
\address{Department of Physics, Massachusetts Institute of
Technology, Cambridge MA 02139\\ and 
CEA, Service de Physique Th\'eorique, 91191 Gif-sur-Yvette
Cedex, France}
\author{James J. Chen\footnote[3]{Current address: Lehman Brothers
Inc., 3 World Financial Center, New York NY 10285.}}
\address{Department of Physics, Massachusetts Institute of
Technology, Cambridge MA 02139}

\maketitle

\begin{abstract}
We report results on  the temperature dependence of the susceptibilities of a set of MBE-grown short-period
EuTe/PbTe antiferromagnetic superlattices having different EuTe layer thicknesses.
In-plane and orthogonal susceptibilities have been measured and display a strong anisotropy at low
temperature, confirming the occurrence of a magnetic phase transition in the thicker samples,
as seen also in neutron diffraction studies. We suggest that dipolar interactions  stabilize
antiferromagnetic long-range order in an  otherwise isotropic system and we present numerical and
analytical results for the low-temperature orthogonal susceptibility.

\end{abstract}



\section{Introduction}
\label{s1}

Magnetic films and multilayers have been a subject of intense study
since they provide experimental realizations for various two-dimensional (2D)
magnetic models.\cite{farrow93,bland94} Most of the recent literature has focused
on {\it{metallic}} magnetic structures. However, {\it{insulating}} antiferromagnetic structures
provide an opportunity to study magnetic long-range order in 2D layered systems of {\it{localized}}
spins. Among those, the EuTe/PbTe superlattice (SL) structures are of special interest since
only one of the two components, EuTe, is magnetic.

Bulk EuTe is a type-II antiferromagnet of the family of Europium chalcogenides, with the structure of
NaCl. Its magnetic moments
arise from the strongly localized $4f$ electrons of the Eu$^{2+}$ atoms which are in a 
symmetric $^8S_{7/2}$ ground state. Thus, the europium chalcogenides have long been considered
ideal realizations of isotropic Heisenberg models.\cite{Wachter79,Mauger86}
Antiferromagnetic resonance experiments have confirmed that, once dipolar interactions are taken into
account, the residual anisotropy in EuTe is negligible.\cite{Streit80,Battles70}
Bulk EuTe has a N\'eel temperature ($T_N$) of 9.8 K. Below $T_N$, spins belonging to a single (111) plane
are parallel but antiparallel to spins in adjacent (111) planes. The magnetic properties of EuTe are
described by a Heisenberg Hamiltonian with two exchange constants: $J_1$ (nearest neighbors, ferromagnetic) and $J_2$
(next-nearest neighbors, antiferromagnetic): 
\begin{equation}
{\cal H}_{ex} = J_1\sum_{nn} {\bf S}_i \cdot {\bf S}_j  + J_2\sum_{nnn} {\bf S}_i \cdot {\bf S}_j \ \ .
\label{e1}
\end{equation}
The ${\bf S}_i$ vectors denote Eu$^{2+}$ spins, which have magnitude 7/2.
$J_1$ and $J_2$ are not known very precisely; currently accepted values are:
$J_1/k_{B} = -0.04 \pm 0.01 K$ and $J_2/k_{B}  = 0.15 \pm 0.01 K$, where $k_{B}$ is the Boltzmann constant.\cite{Wachter79,Zinn76}

Although the Hamiltonian in eq. (\ref{e1}) has full rotational symmetry, neutron diffraction experiments
show that the spins lie in (111) planes.\cite{Will63} This easy-plane anisotropy is
properly accounted for by
adding dipole-dipole interactions to the exchange Hamiltonian in eq. (\ref{e1}).\cite{Kaplan54,Keffer57}

In this article we present experimental and theoretical studies of the susceptibilities of
short-period EuTe/PbTe SL's. We use the notation EuTe($\xi$)/PbTe($\eta$) to denote a SL structure
whose period
consists of $\xi$ monolayers of EuTe and  $\eta$ monolayers of PbTe. Each sample used in the present
work consisted of 400 such periods and was prepared so that $\eta = 3\xi$. Since samples are grown in the (111)
direction, the structure within the EuTe monolayers is that of a triangular lattice and
the monolayers are stacked
according to the ABC sequence.
Section II describes our samples and presents results for their temperature-dependent zero-field susceptibilities.
Section III presents a discussion of the experimental results, then Section IV presents
a mean-field determination of the order parameter, a Monte-Carlo
simulation of the susceptibilities of an EuTe(3)/PbTe(9) sample, and a calculation of the out-of-plane
susceptibility at low temperature.
Section V summarizes our findings.

\section{Experimental results}
\subsection{Sample Characterization}

Experiments were conducted on SL samples of EuTe($\xi$)/PbTe($3\xi$) for $1\leq\xi\leq 7$.
In each sample the SL stack was grown on a 3000 \AA~ PbTe (111) oriented  buffer
layer, itself grown on a BaF$_2$ (111) substrate. A  500 \AA~ PbTe cap layer was used to prevent
oxidation of the highly reactive EuTe. Details
of the MBE growth process have been published elsewhere.\cite{Gunther93,Gunther95,Gunther94}
The SL's have approximately square wave
composition modulation, as reflected by the multiple narrow SL peaks
of the high-resolution x-ray diffraction data.\cite{chen96}
Electron spin resonance
(ESR) experiments show very little interdiffusion 
at the EuTe-PbTe interface.\cite{willamowski95} 
Furthermore, careful {\it in situ} scanning tunneling microscopy (STM)
investigations have shown that the
PbTe and EuTe heterointerfaces are quite smooth
on a length scale of at least 200 \AA, with imperfections
strictly limited to single monolayer steps.\cite{Frank94}
The {\it ex situ} cross-sectional TEM images also exhibit
smooth PbTe and EuTe interfaces.\cite{Shima95}  

The magnetic properties of the SL's used  have
been studied previously by SQUID magnetometry,\cite{chen96} and elastic neutron
scattering.\cite{Giebultowicz95}  Magnetic hysteresis curves and neutron diffraction
spectra taken at 1.8K and 4.2K  show that for all samples with $\xi \geq 3$
a transition to a low-temperature ordered phase takes place at $T_N \geq$ 4.2K, the order being
that of a type II antiferromagnet, i.e.
identical to that of bulk EuTe. Since the MBE samples are grown along the (111) direction,
this implies that the spins in each EuTe monolayer order ferromagnetically and are antiparallel
to those in neighboring monolayers. Moreover, the spins lie within the EuTe monolayers. 
Static magnetization measurements taken parallel
to the SL plane show no detectable in-plane anisotropy.\cite{Thesis}

\subsection{Experimental Conditions}
\begin{figure}
\epsfysize=5.cm{\centerline{\epsfbox{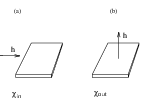}}}
\caption{The two principal orientations of the sample with respect
to the AC probing field for the $\chi_{in}$ (a), and $\chi_{out}$ (b)
susceptibility measurements.}
\label{f01}
\end{figure}

Our susceptiblility measurements used a Quantum Design (MPMS5) AC susceptometer.
Susceptibilities were measured in two geometries :  $\chi_{in}$ was measured with 
the AC probing field $h$ parallel to the SL plane, $\chi_{out}$ was measured
with $h$ orthogonal to the SL plane (see Fig. \ref{f01}).

For each sample, $\chi_{in}$ and $\chi_{out}$ have been measured
as a function of temperature from 15 K in the paramagnetic
region to below the transition temperature, 
in a nominal zero external static magnetic field (i.e., less than 2
Gauss) with a 20 Hz
AC probing field of 4 Gauss. Since the probing field is small,
diamagnetic contributions from the BaF$_2$ substrate and the PbTe buffer layer can
safely be ignored. 
A study of the dependence of the susceptibility on $h$ and on the AC frequency
indicates that our measurements are always in the linear regime of the 
static susceptibility.

\subsection{Experimental Susceptibilities}
\begin{figure}
\hskip -3.truecm \epsfxsize=15.truecm{\centerline{\epsfbox{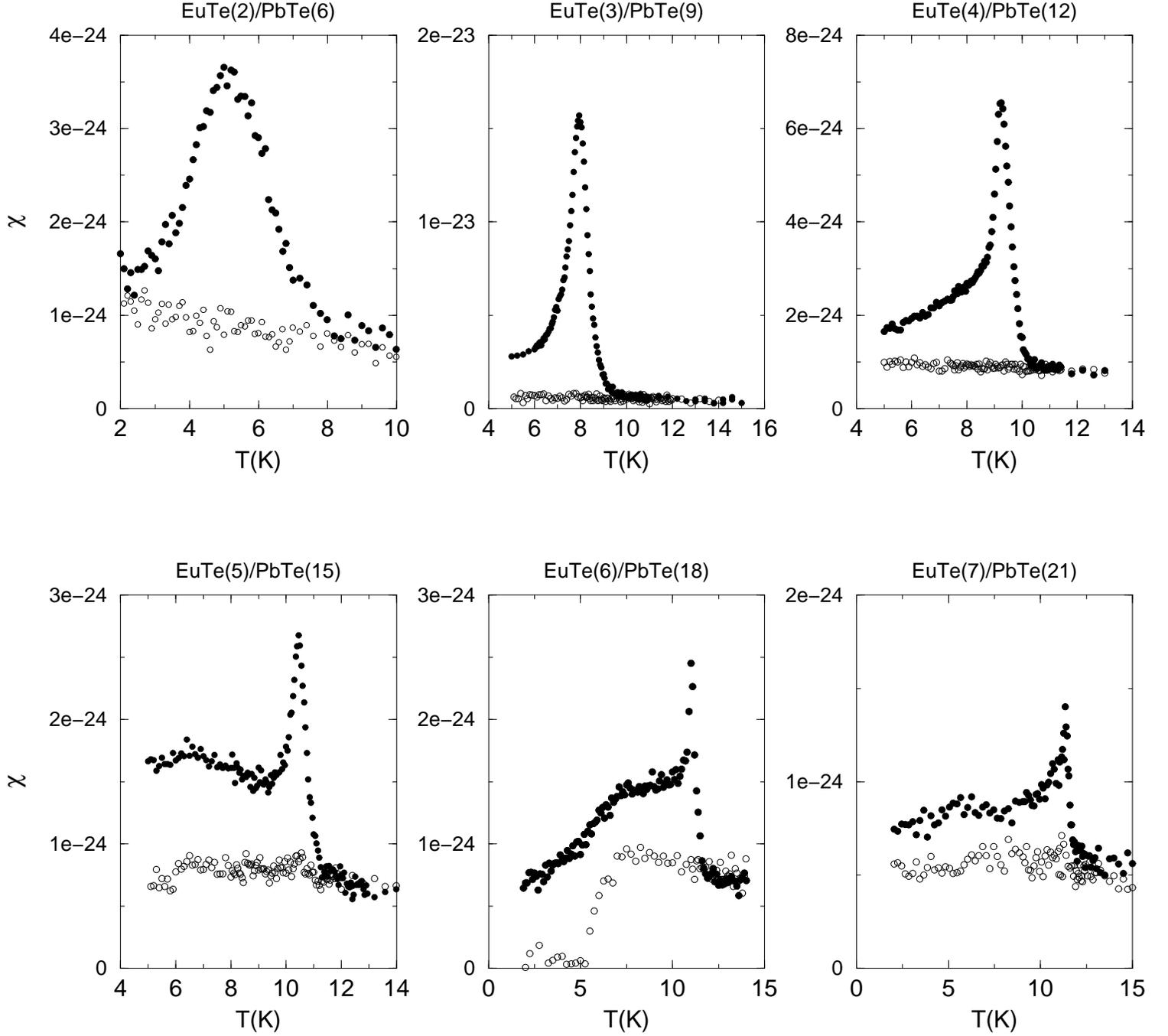}}}
\vskip 1.2truecm
\caption{$\chi_{in}$ (in-plane, $\bullet$) and $\chi_{out}$ (orthogonal, $\circ$) susceptibilities in emu 
normalized per Eu atom, for samples EuTe(2)/PbTe(6) through EuTe(7)PbTe(21).}
\vskip 0.5truecm
\label{totalgraph}
\end{figure}

Figure \ref{totalgraph} displays the temperature dependence of the in-plane and ou-of-plane magnetic
susceptibilities $\chi_{in}$ and  $\chi_{out}$ for
samples EuTe(2)/PbTe(6) through EuTe(7)/PbTe(21).
The experimental susceptibilities have been normalized by the
number of Eu$^{2+}$ spins, determined using high temperature susceptibility data; no demagnetizing corrections
have been applied to the data. In the following we will be mostly interested in $\chi_{out}$, whose magnitude
changes little below the temperature at which $\chi_{out}$ and $\chi_{in}$ separate. Since there is no
observable difference in $\chi_{out}$ and $\chi_{in}$ above this temperature we can assume that demagnetizing
corrections are not significant. 

\begin{figure}
\epsfxsize=15.truecm{\centerline{\epsfbox{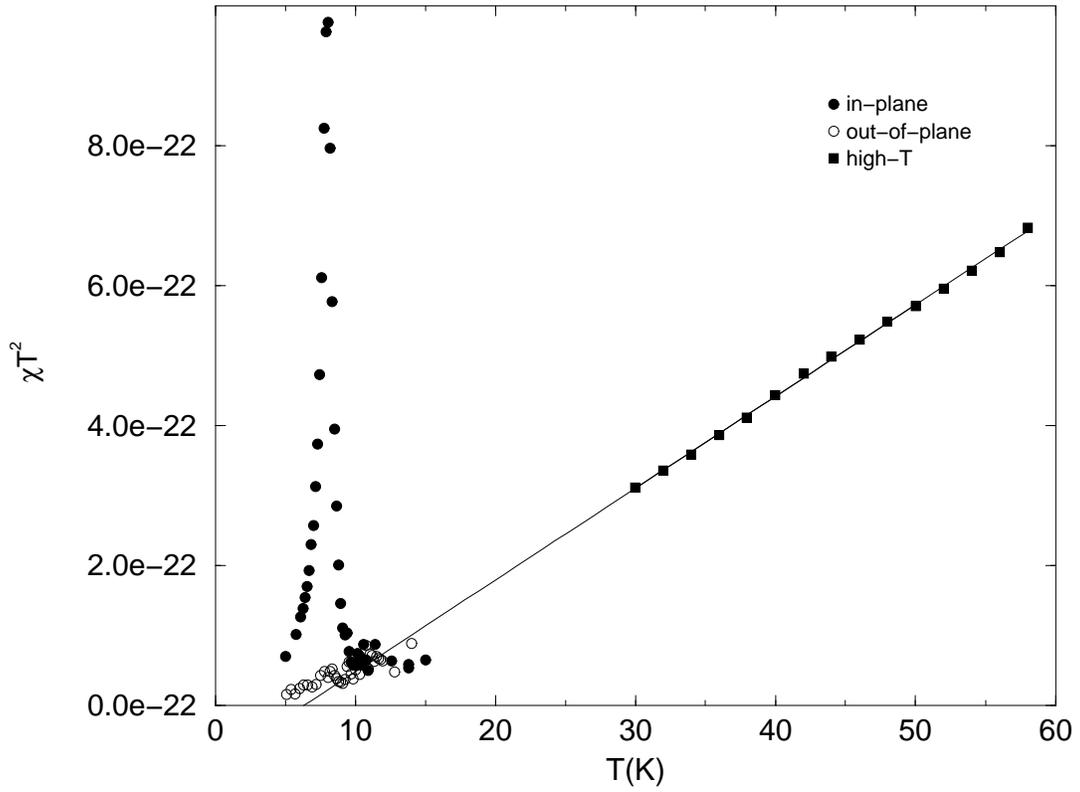}}}
\caption{A plot of $\chi T^2$ for sample EuTe(3)/PbTe(9) in the paramagnetic phase (squares)
and in the ordered phase, in the in-plane ($\bullet$) and out-of-plane ($\circ$) directions, in emu
normalized per Eu atom.}
\label{chit2}
\end{figure}

For all samples with $\xi > 1$ a plot of $\chi T^2$ versus $T$
showed that, as the temperature is lowered, the Curie
behavior of $\chi_{in}$ and $\chi_{out}$ persists until they separate at a temperature $T_s$.\cite{Thesis}
Fig. \ref{chit2} shows such a plot for sample EuTe(3)PbTe(9). 
Generically, below the point of separation a very anisotropic behavior of the susceptibility is observed,
with $\chi_{out} \leq \chi_{in}$, an unusual feature. Furthermore,  while  $\chi_{in}$ displays a peak,
$\chi_{out}$ has a very mild temperature dependence below $T_s$.
In addition, the magnitude of $\chi_{out}$ at low temperature, of the order of
$1 . 10^{-24}$ emu/spin, has little dependence on the thickness of the EuTe layer. 
Since we do not have other experimental data  (e.g. specific heat) that would allow
us to define more precisely the transition temperature $T_N$, we assumed  that it lies in between $T_s$
and the temperature at which $\chi_{in}$ is
maximum. $T_N$ increases with the thickness of the EuTe layer, as expected, and reaches values
higher than that for bulk EuTe for samples with $\xi \ge 5$, an unexpected result. A possible explanation
could be that coupling constants have values different from the bulk ones, a point developed
further in section \ref{discuss}. 
We comment now on samples that show non-generic behavior.
\subsubsection{EuTe(1)/PbTe(3)}
\begin{figure}
\epsfxsize=15.truecm{\centerline{\epsfbox{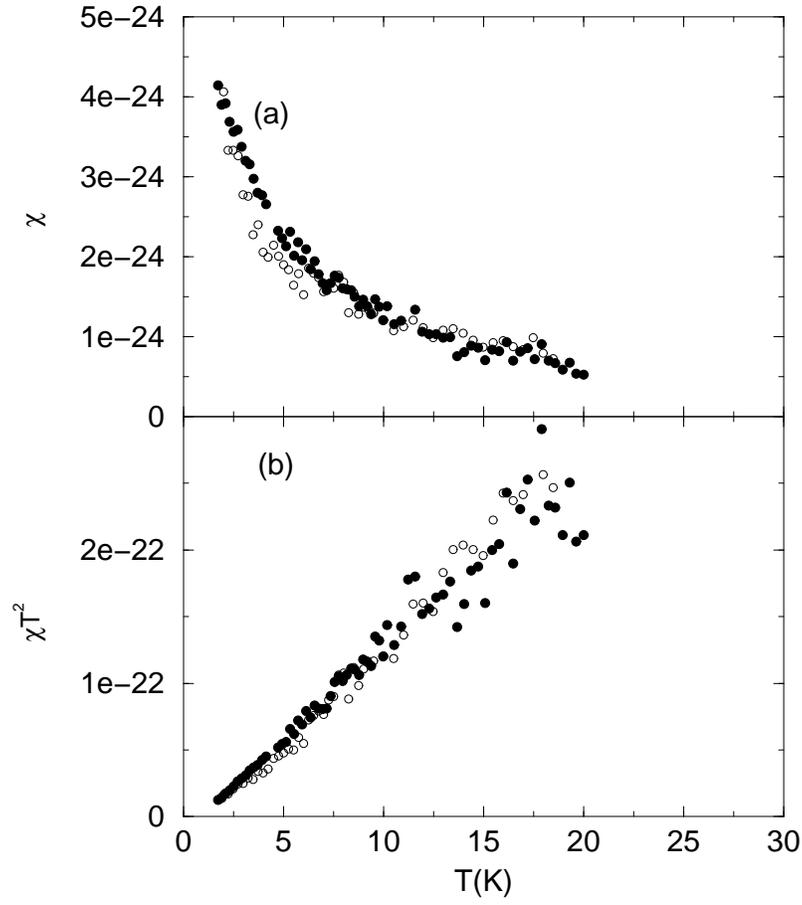}}}
\caption{(a) $\chi_{in}$ (in-plane, $\bullet$) and $\chi_{out}$ (orthogonal, $\circ$) susceptibilities in emu
normalized per Eu atom, for sample EuTe(1)PbTe(3). 
(b) Same susceptibilities, multiplied by $T^2$.}
\vskip 0.5truecm
\label{sample13}
\end{figure}

$\chi_{in}$ and $\chi_{out}$ as well as $\chi_{in}T^2$ and $\chi_{out}T^2$ have been measured
for $T > 1.8$K. They are plotted in Fig. \ref{sample13}. One sees that in this temperature regime
sample EuTe(1)/PbTe(3) remains in a paramagnetic phase : $\chi_{in}$ and $\chi_{out}$
coincide and exhibit a Curie behavior. The absence of a transition in this temperature range can be
understood
by noting that in a monolayer geometry the $J_2$ exchange coupling is not present
and the energy scale is set by the nearest-neighbor exchange
coupling only, which is very small. We expect however a transition at a lower temperature. 
\subsubsection{EuTe(2)/PbTe(6)}
Elastic neutron scattering spectra show no long-range order at 4.2K but they do show a peak corresponding
to type II antiferromagnetic ordering at 1.8K. Thus the broad maximum in $\chi_{in}$ does not signal a 
transition from the paramagnetic to the antiferromagnetic phase. 
\subsubsection{EuTe(6)/PbTe(18)}
The sharp drop in $\chi_{out}$ around 5K is reproducible. However we only had one EuTe(6)/PbTe(18) sample.
EuTe(6)/PbTe(18) is the only sample in our series that displays this feature, as yet unexplained.

\section{Discussion}
\label{discussion}

A qualitatively
different behavior for $\chi_{in}$ is  expected {\it a priori}, depending on whether the
number of EuTe monolayers per SL period is odd or even. In the first case, each period should behave as a 
ferromagnet and there should be a peak
in $\chi_{in}$ whereas in the latter case each period should behave as an antiferromagnet and
we should see a smooth maximum.
Although we use integers $\xi
$ and $\eta$ to label our samples, the average
thickness of, say, the EuTe layer in an actual SL period as determined by X-ray scattering is
fractional because of the interface structure
so that the odd/even effect is in fact expected to be blurred.\cite{chen96}

In our view, the main issues raised by our data is the existence of a phase transition in samples
with $\xi \geq 3$ at temperatures comparable to the bulk T$_N$. As mentioned above, neutron
diffraction spectra unambiguously demonstrate that these samples have a low-temperature ordered
phase. Also, anisotropy in the Hamiltonian for Eu$^{2+}$ spins is negligible, so that our samples
can be considered as representing a model 2D Heisenberg system. In such a system, with isotropic exchange couplings,
the transition should occur at T=0K.\cite{Mermin66} The Hamiltonian is however incomplete and we need to supplement
the exchange couplings with dipole-dipole interactions. Although the dipole-dipole coupling may be
weak, it  breaks the rotational symmetry and is  long-range, which prevents application
of the Mermin-Wagner theorem.\cite{Mermin66} Dipolar couplings are known to
have large effects in metallic thin
films, in which they compete with uniaxial anisotropy.\cite{Pescia90,Erickson912}

Early work \cite{Maleev76,Pokrovski77} has demonstrated the 
possibility of a phase transition driven by dipolar interactions in a 2D isotropic spin
system. More recent work has investigated the dependence of $T_N$ on the magnitude of the dipolar
coupling.\cite{Chui94,Pich,Arruda95}. It is our
hypothesis that dipolar interactions are responsible for stabilizing magnetic long-range order in
EuTe/PbTe SL's. In the next section we discuss some implications of this hypothesis.

The observation that $\chi_{out} \leq \chi_{in}$ for all samples can be understood by noting that
there are actually three pertinent susceptibilities in our system :
$\chi_\parallel$, along the direction of the order parameter in the SL plane, $\chi_\perp$, in
the SL plane, but orthogonal to the direction of the order parameter, and $\chi_{out}$, orthogonal
to the SL plane. In our case, $\chi_\perp$ and $\chi_{out}$ will be
different because of dipolar interactions, and $\chi_{out}$ will be smaller than $\chi_\perp$. Usually
one has $\chi_\parallel \leq \chi_\perp$, and in our case we expect
$\chi_\parallel \leq \chi_{out} \leq \chi_\perp$. The assumption that in each atomic layer the spins
belong to domains with random orientations yields : $\chi_{in}={1\over 2}(\chi_\parallel +
\chi_\perp)$. It is thus possible for $\chi_{out}$ to be smaller than $\chi_{in}$.

\section{Theory}
\label{s3}

In this section, we use the following Hamiltonian to describe the interaction of Eu$^{2+}$ spins:

\begin{equation}
{\cal H} = J_1\sum_{nn} {\bf S}_i \cdot {\bf S}_j  + J_2\sum_{nnn} {\bf S}_i \cdot {\bf S}_j  
+ \sum_{ij, \alpha\beta} Q_{\alpha\beta}({\bf r}_j-{\bf r}_i)
S_i^\alpha S_j^\beta ,
\label{e2}
\end{equation}
where the third sum runs over all sites $i$ and $j$ in the SL and on spin components $\alpha$, $\beta$.
Here $Q_{\alpha\beta}({\bf r}_j-{\bf r}_i)$ is the dipolar tensor which reads: 
\begin{equation}
Q_{\alpha\beta}({\bf r}_j-{\bf r}_i) = \frac{(g\mu_b)^2}{2}\left( \frac{\delta^{\alpha\beta}}{r_{ij}^3}
-3\frac{r_{ij}^\alpha r_{ij}^\beta}{r_{ij}^5} \right)
\ .
\end{equation}
In this expression, $\mu_b$ is the Bohr magneton and $g$ is the Land\'e $g$ factor which we will take equal
to 2. 
Throughout this section we will approximate the spins ${\bf S}_i$ by {\it classical} vectors.
This is justified by the large magnitude $S=\frac{7}{2}$ of the Eu$^{2+}$ spins and the fact that we don't
expect quantum effects in the temperature range we will be considering. 
The structure within the EuTe layers is that of a triangular lattice and the layers are stacked according
to the ABC sequence. We will take the $x$ and $y$ axes in the plane of the layer and the  $z$ axis orthogonal
to the layer plane. 

\subsection{Orders of magnitude}

Let us denote by $E_\parallel$ and $E_\perp$ 
the dipolar energies per spin of an EuTe monolayer, assuming the spins 
are ferromagnetically aligned either in the layer plane 
or orthogonal to the layer plane. $E_\perp$ is given by :

\begin{equation}
E_\perp = \sum_i Q_{zz}({\bf r}_i) \ = \ \frac{(g\mu_bS)^2}{2} \sum_i {1\over r_i^3}
\label{e4}
\ ,
\end{equation}
where the sums run on the sites of a single layer.
$E_\parallel$ is given by :

\begin{equation}
E_\parallel = \sum_i Q_{xx}({\bf r}_i) \ = -\frac{(g\mu_bS)^2}{4} \sum_i {1 \over r_i^3}
\ ,
\end{equation}
where we have used the fact that $Q$ is traceless and rotationally invariant in the $xy$ plane. 
Notice that $E_\parallel < 0$, which favors in-plane alignment, as observed in neutron diffraction experiments.
Using the value of $\sum_i {1\over r_i^3} = 11.035/a^3 $ for a triangular lattice
and the value of the in-plane lattice parameter $a=4.6$\AA \  taken from X-ray data, we get
$E_\parallel = -0.86$K and $E_\perp = 1.72$K, to be compared with the exchange energies $J_1S^2 = -0.5$K,
$J_2S^2 = 1.8$K
and the  average exchange energy per spin $E_{ex}= -6J_2S^2 = 11$K,
which we have estimated using bulk EuTe values for $J_1$,$J_2$.

Next we estimate the size of the interlayer dipolar energies.
To this end we have used Ewald summation techniques that allow one to rewrite $1\over r^3$
sums as fast-converging series.\cite{Maleev76,Benson69} Let us consider two neighboring EuTe
monolayers (1) and (2) a distance $h$ apart. 
Let ${\bf r}$ be the vector joining a lattice site in layer (1) to a lattice site in layer (2). 
Assuming that the spins in layer (1) and (2)  are all ferromagnetically aligned but with opposite directions
depending on which layer they belong to,  the interaction energy of a spin in layer (1) with all spins in
layer (2) is:
\begin{equation}
E = -{(g\mu_bS)^2\over 2 }\frac{2\pi}{\cal A} \ \sum_{\bf G} \frac{G_x^2}{G}e^{-hG}\cos({\bf G}.{\bf r})
\end{equation}
where the sum runs over all reciprocal lattice vectors $\bf G$, and $G$ and $G_x$ are, respectively, the modulus
and the $x$-component of $\bf G$ and $\cal A$ is the area of the triangular lattice unit cell in the layer.
We thus get :
$$
E = 0.1014 \ \frac{(g\mu_bS)^2}{a^3} \ = \ 0.0317 {\rm K} \ .
$$
This energy is much smaller than the intralayer energy $E_\parallel$.
If the spin density in the layer
were uniform, no field would be created outside the layer and this energy would be 0.
The very existence of a lattice
structure
within the layer makes it finite. Furthermore, the interaction energy is expected to decay fast as the
distance from the layer 
becomes larger than the in-plane lattice constant. For instance,
at a distance two layers away, this energy is $-8.8 \ 10^{-5}$K, and three layers away it is $4.7 \ 10^{-7}$K.

As a result,  we can safely discard all interlayer couplings as well as interperiod couplings, which
couple spins belonging to different SL periods. Although small, the intralayer coupling has to be retained
for the reasons mentioned in section \ref{discussion}.


\subsection{Mean Field analysis}
\label{meanfield}
We now turn to a Mean Field treatment of our problem. Our aim here is to identify
the order parameter for the phase transition rather than find the expression for $T_N$.. We consider
a single period of a SL which consists of $N$ EuTe monolayers. The spins are labelled with two indices:
$i$ denotes their position in a layer, $n$ the layer to which they belong. 
We rewrite the Hamiltonian as:
\begin{equation}
{\cal H} = \sum_{ijnm,\alpha\beta} H_{in\alpha, jm\beta} \ S_{in}^\alpha S_{jm}^\beta
\label{fullham}
\end{equation}
where the matrix $H$ is defined by:
\begin{equation}
H_{in\alpha, jm\beta} = \delta^{\alpha\beta} J({\bf r}_j-{\bf r}_i, m-n)
+\delta^{nm}Q_{\alpha\beta}({\bf r}_j-{\bf r}_i)
\end{equation}
in which J denotes the matrix of the exchange couplings and its elements are equal to $J_1$ if ($in$) and ($jm$)
are nearest neighbors, to $J_2$ if they are next-nearest neighbors,  and to zero otherwise.
The $\delta^{nm}$ factor expresses the fact that interlayer dipolar couplings are neglected.

A mean-field calculation in our context amounts to diagonalizing $H$ and finding its lowest
eigenvalue. The magnitude of the latter determines $T_N$ while the associated eigenvector defines the 
order parameter for the transition. \cite{Orland}

Since $Q$ is diagonal in the layer indices we concentrate first on $J$. 
We define in-plane Fourier transforms ${\bf S}_{{\bf q}n}$ for the spins through:
\begin{equation}
{\bf S}_{{\bf q}n} = {1\over{\sqrt {\cal N}}} \ \sum_i \ {\bf S}_{in} e^{i{\bf q}.{\bf r}_i}
\label{Fourier1}
\end{equation}
where $\cal N$ is the number of spins per layer. Likewise we define $J({\bf q},m-n)$ as:
\begin{equation}
J({\bf q},m-n) = \sum_{\bf d} J({\bf d},m-n) e^{i{\bf q}.{\bf d}} \ .
\label{Fourier2}
\end{equation}
Since we expect the ordered phase to be homogeneous in the plane of the layers, we now restrict ourselves
to the ${\bf q}={\bf 0}$ sector of the Hamiltonian. 
The neighbors and nearest-neighbors of a spin in layer $n$ all belong to layers $n-1, n, n+1$. The only
non-zero matrix elements of $J({\bf 0}, m-n)$ are thus:
$$
J({\bf 0},0)=6J_1 \ \ \mbox{and} \ \  J({\bf 0},1) = J({\bf 0},-1) = 3(J_1+J_2).
$$
The ${\bf q}={\bf 0}$ part of the Hamiltonian now reads:
\begin{equation}
{\cal H}_{{\bf q}={\bf 0}} = \sum_{nm} A_{nm} \ {\bf S}_{{\bf 0}n} {\bf S}_{{\bf 0}m}
\label{Adef}
\end{equation}
where $A_{nm}=J({\bf 0},m-n)$. Now $A$ can be diagonalized in the basis of $N$-dimensional orthonormal
vectors $T_k$,  whose components are:
\begin{equation}
(T_k)_n = \sqrt{2\over{N+1}} \sin\left(\frac{nk\pi}{N+1}\right)
\label{tkdef}
\end{equation}
where $k$ is an integer ranging from 1 to $N$.
The corresponding eigenvalues are:
\begin{equation}
a_k= 6J_1 + 6(J_1+J_2)\cos\left(\frac{k\pi}{N+1}\right)
\ .
\end{equation}

In the same way as for $J$ in eq. (\ref{Fourier2}) one can define a Fourier
transform $Q_{\alpha\beta}({\bf q})$. It is diagonal
for ${\bf q}={\bf 0}$ with: 
\begin{eqnarray}
Q_{xx}({\bf 0}) = Q_{yy}({\bf 0}) &=& -{\cal C} \label{Qdiag1} \\
Q_{zz}({\bf 0}) &=& 2{\cal C}
\label{Qdiag2}
\end{eqnarray}
where
\begin{equation}
{\cal C}=\frac{(g\mu_b)^2}{4}\sum_i{1\over r_i^3}.
\label{defineC}
\end{equation}
The lowest eigenvalue of matrix $H$ is thus $a_N-2{\cal C}$ and the mean field $T_N$ is given by:
\begin{eqnarray}
T_N &=& -\frac{2S^2}{3k_{B}}(a_N-2{\cal C}) \nonumber \\
 &=& \frac{2S^2}{3k_{B}}\left(-6J_1 -6(J_1+J_2)\cos\left(\frac{N\pi}{N+1}\right)
+2{\cal C}\right) \ .
\end{eqnarray}
Using bulk values for $J_1$, $J_2$ we find $T_N = 6.93$K.
The associated order parameter is a linear combination of the in-plane projections of the spins with
weights $(T_k)_n$ defined in (\ref{tkdef}):
\begin{equation}
{\bf M}_N = \sqrt{2\over{N+1}} \sum_n \sin\left(\frac{nN\pi}{N+1}\right) {\cal S}_n
\label{Mndef}
\end{equation}
where ${\cal S}_n$ denotes the in-plane projection of the sum of all spins belonging to layer $n$.
Note that, as $N\rightarrow\infty$ the usual antiferromagnetic staggered
magnetization is recovered. We will in the next section check that ${\bf M}_N$ is
indeed the correct order parameter.

\subsection{Monte Carlo simulation}
\label{Montecarlo}

We have performed a Monte Carlo simulation for our system in order to check that dipolar
interactions can drive a transition at a temperature $T_N > 0$ and
can generate an anisotropy in the susceptibility similar to that observed in experiments.
We have also checked  the relevance of the mean field order parameter. 

\begin{figure}
\epsfxsize=15.truecm{\centerline{\epsfbox{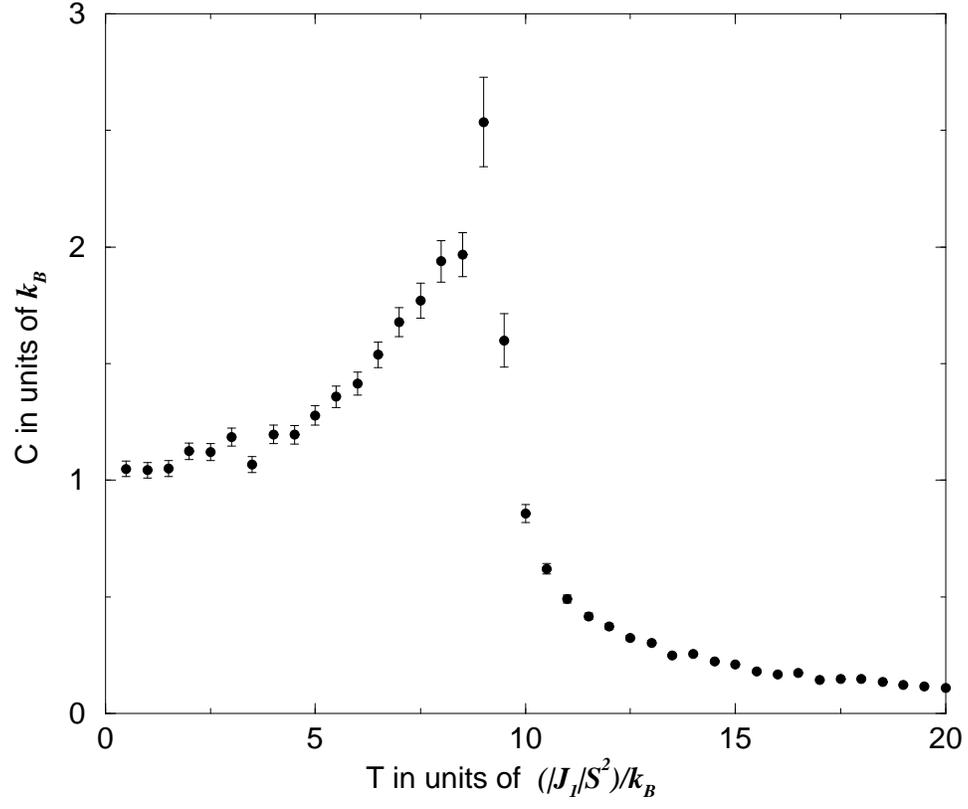}}}
\caption{Simulated specific heat for a EuTe(3)/PbTe(9) system in units of $k_B$.}
\label{spheat}
\end{figure}

We present here results of a Monte Carlo simulation carried out on a system of 3 layers,
each consisting of 23$\times$23 spins, with periodic boundary conditions in the plane of the layers, 
in order not to introduce
in-plane anisotropy. Each layer is thus mapped to a torus.  The full Hamiltonian (\ref{fullham}) has been used, where the distance
$r_{ij}$ between pairs of sites has been taken to be the smallest distance on the 
torus between sites $i$ and $j$. All couplings in the Hamiltonian have been expressed in units of $J_1$
and we have used EuTe bulk values for the ratios $J_2/J_1 = -3.75$ and
$(g\mu_b)^2/(J_1a^3) = 0.64$, which are the only parameters of our model. 
The heat-bath algorithm with sequential updating of the spins has been used, with 400 equilibration sweeps and 2000
sweeps with a measurement after each sweep. Error bars have been carefully computed as standard deviations of 
estimators of the observables. Three susceptibilities have been evaluated: $\chi_z$, along the normal
to the layers, corresponding to $\chi_{out}$,  and $\chi_x$ and $\chi_y$ in the plane of the layers.
The specific heat has also been measured to check that it has a limit of $k_{B}$ per spin as
$T\rightarrow 0$, a
general property of classical spin systems.\cite{fisher64} It is shown in Fig. \ref{spheat} for the
EuTe(3)/PbTe(9) system. Susceptibilities have been computed as:
$$
\chi_\alpha = \frac{(g\mu_b)^2}{N{\cal N}k_{B}T}\left( \langle S_\alpha^2\rangle -
\langle S_\alpha\rangle^2 \right)
$$
where $\bf{S}$ is the total spin and ${\cal N}$ is the number of spins per layer.

Because the simulation has been done on a finite system, the in-plane rotational symmetry is not broken in the
ordered phase with the result that the above averages are not well defined. We have thus evaluated
$\chi_x$ and $\chi_y$ only at temperatures higher than the transition temperature. 

\begin{figure}
\epsfxsize=16.truecm{\centerline{\epsfbox{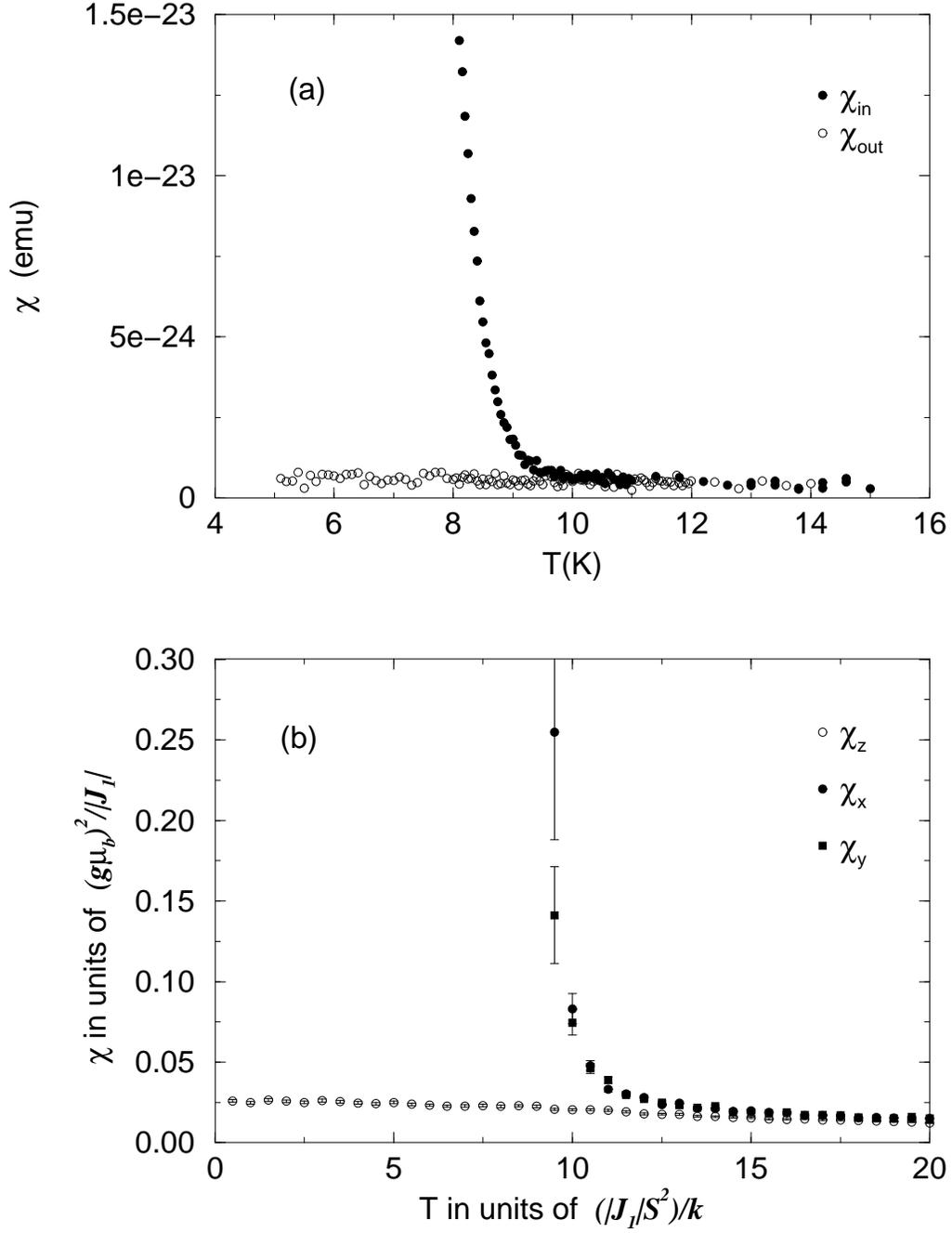}}}
\caption{(a) Measured $\chi_{in}$ and $\chi_{out}$ for the EuTe(3)/PbTe(9) sample.
(b) Simulated $x, y, z$ susceptibilities for a 3-layer system of size 23$\times$23.}
\vskip 0.5truecm
\label{Chime+si}
\end{figure}

Experimental and simulated susceptibilities have been plotted as a function of temperature
in Fig. \ref{Chime+si}. One can see that the simulation qualitatively reproduces the anisotropy observed in
the experiments, with a remarkable flatness of $\chi_z$ ($\chi_{out}$) at low temperature. We have also run
simulations on 3-layer systems of sizes 29$\times$29 and 13$\times$13 and these two systems
didn't show any significant difference in the transition temperature or the low-temperature magnitude of
$\chi_z$.

\begin{figure}
\epsfxsize=16.truecm{\centerline{\epsfbox{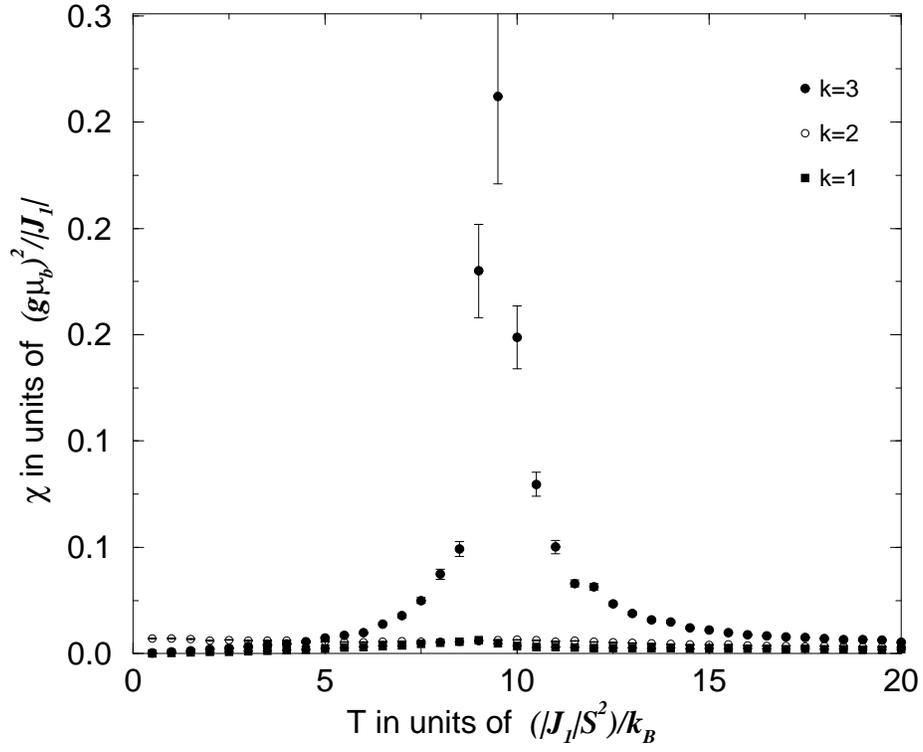}}}
\caption{Simulated susceptibilities corresponding to order parameters ${\bf M}_k$, for a 3-layer system
of size 23$\times$23.}
\vskip 0.5truecm
\label{Chipar123}
\end{figure}

We then checked the relevance of the order parameter ${\bf M}_N$ found in the mean field approach of section
B. In the same way as ${\bf M}_N$ in eq. (\ref{Mndef}) has been defined using vector $T_{k=N}$ each vector $T_k$
can be used to build an order parameter ${\bf M}_k$, defined by:
\begin{equation}
{\bf M}_k=\sqrt{2\over{N+1}}\sum_n \sin\left( \frac{nk\pi}{N+1}\right) {\cal S}_n
\end{equation}
using the same notations as before. The ${\bf M}_k$'s are linear combinations of the in-plane projection of
the spins. One could similarly define linear combinations of their $z$ components, although we know that spins
order in-plane. For a 3-layer system, three order parameters can be defined: 

\begin{eqnarray*}
{\bf M}_1 & = &\frac{1}{\sqrt{2}}\left( \frac{1}{\sqrt{2}}{\cal S}_1 + {\cal S}_2 + \frac{1}{\sqrt{2}}
{\cal S}_3 \right) \\
{\bf M}_2 & = & \frac{1}{\sqrt{2}}\left( {\cal S}_1 - {\cal S}_3 \right) \\
{\bf M}_3 & = & \frac{1}{\sqrt{2}}\left( \frac{1}{\sqrt{2}}{\cal S}_1 - {\cal S}_2 + \frac{1}{\sqrt{2}}
{\cal S}_3 \right) \ .
\end{eqnarray*}
One can define susceptibilities $\chi_k$ for the moduli of these three order parameters $k=1, 2, 3$ as: 
$$
\chi_k = \frac{1}{N{\cal N}k_BT}\left( \langle{\bf M}_k^2\rangle -\langle |{\bf M}_k |\rangle^2 \right) \ .
$$
Because vectors $T_k$ are normalized, the $\chi_k$ all have
the same leading behavior at high temperature. These susceptibilities are plotted in Fig.\ref{Chipar123}.
$\chi_3$ is the susceptibility that displays a sharp peak, thus suggesting that ${\bf M}_3$ is the
appropriate order parameter for describing the phase transition. 

\subsection{Low-temperature orthogonal susceptibility}

\begin{figure}
\epsfxsize=12.truecm{\centerline{\epsfbox{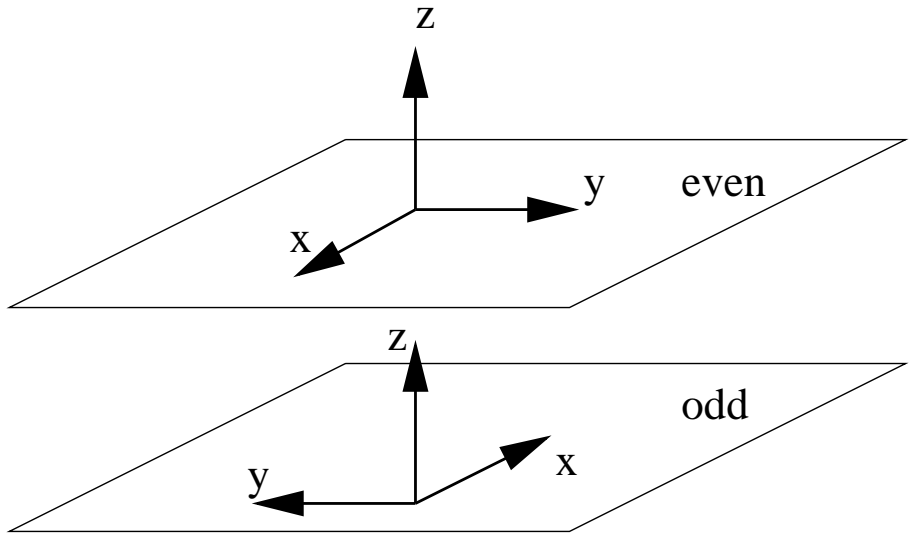}}}
\caption{Layer-dependent reference frames as defined in section IV. D.}
\vskip 0.5truecm
\label{frame}
\end{figure}

Since $\chi_{out}$ is flat at low temperature, it is desirable to have an estimate of
its magnitude. We present here a calculation of $\chi_{out}$ using the spin Hamiltonian given
in eq. (\ref{fullham}).
$$
{\cal H} ={\cal H}_{ex} + {\cal H}_d
$$
where the exchange term is:
\begin{equation}
{\cal H}_{ex} = \sum_{ij,nm}J({\bf r}_j-{\bf r}_i,m-n) \ {\bf S}_{in} \cdot {\bf S}_{jm}
\end{equation}
and the dipolar term is:
\begin{equation}
{\cal H}_d = \sum_{ij,n} Q_{\alpha\beta}({\bf r}_j-{\bf r}_i) \ S_{in}^\alpha S_{jn}^\beta \ .
\end{equation}
In each layer we define a frame of reference as shown in Fig. \ref{frame}, such that the $y$ axis lies along
the direction of the layer magnetization, while the direction of the $z$ axis, orthogonal to the layers, is the
same for all layers. The spin components $\sigma_{in}^\alpha$ in the layer-dependent frames are related to the
$S_{in}^\alpha$ through:

\begin{eqnarray}
S_{in}^x & = & (-1)^n \sigma_{in}^x \nonumber \\
S_{in}^y & = & (-1)^n \sigma_{in}^y \\
S_{in}^z & = & \sigma_{in}^z \nonumber \ . 
\end{eqnarray}
At low temperature $\sigma_{in}^z$ and $\sigma_{in}^x$ will  be small, while $\sigma_{in}^y$
will be finite, with a fixed sign. $\chi_{out}$ is defined as:
$$
\chi_{out} = \frac{(g\mu_b)^2}{N{\cal N}k_{B}T}\langle (\sigma^z)^2\rangle
$$ 
where $\sigma^z = \sum_{in}\sigma_{in}^z $. For classical 3D spins $\langle (\sigma^z)^2\rangle$ is given by:
\begin{equation}
\langle (\sigma^z)^2\rangle =  \frac{1}{Z} \int \prod_{in} 
\frac{d\sigma_{in}^z d\sigma_{in}^x}{S \ |\sigma_{in}^y|} \ (\sigma^z)^2 \ e^{-\beta\cal{H}}
\label{chiT}
\end{equation}
where $\beta = \frac{1}{k_{B}T}$. The partition function $Z$ reads:
$$
Z=\int \prod_{in}
\frac{d\sigma_{in}^z d\sigma_{in}^x}{S \ |\sigma_{in}^y|} \ e^{-\beta\cal{H}} \ .
$$
At low temperature $(\sigma_{in}^z)^2$ and $(\sigma_{in}^x)^2$
will be of order $k_{B}T$ if $\chi_{out}$ is to be finite.
We can thus let the integrals run from $-\infty$ to $+ \infty$, instead of $-S$ to $+S$. 
We note that $\sigma_{in}^y$ appears both in the integration measure and $\cal H$ and is given by:
$$
\sigma_{in}^y = \sqrt{S^2 -({\sigma_{in}^z}^2 + {\sigma_{in}^x}^2)} .
$$
We now expand $\sigma_{in}^y$ in powers of $({\sigma_{in}^z}^2 + {\sigma_{in}^x}^2)$
keeping only the lowest order terms. This amounts to an expansion in powers
of the temperature.
The measure becomes $d\sigma_{in}^z d\sigma_{in}^x /S^2$.
In the expression of $\beta{\cal H}$ only the first-order term need be kept.
We are then left with  a
quadratic form for $\sigma_{in}^z$ and $\sigma_{in}^x$
which we need to diagonalize in order 
to calculate $\langle (\sigma^z)^2\rangle$.

Let us first consider ${\cal H}_{ex}$. In our approximation ${\bf S}_{in} \cdot {\bf S}_{jm}$ reads:

\begin{equation}
{\bf S}_{in} \cdot {\bf S}_{jm} = \sigma_{in}^z\sigma_{jm}^z + (-1)^{m-n} \left( \sigma_{in}^x\sigma_{jm}^x
+(S-\frac{{\sigma_{in}^z}^2 +{\sigma_{in}^x}^2}{2S})(S-\frac{{\sigma_{jm}^z}^2 +{\sigma_{jm}^x}^2}{2S})
\right) \ .
\end{equation}

After discarding constant terms, we get the following expression for ${\cal H}_{ex}$

\begin{equation}
{\cal H}_{ex} = \sum_{ij,nm}\left( J({\bf r}_j-{\bf r}_i,m-n) + \alpha_n\delta_{ij}\delta_{nm} \right)
\left( \sigma_{in}^z\sigma_{jm}^z + (-1)^{m-n}\sigma_{in}^x\sigma_{jm}^x \right)
\end{equation}

where $\alpha_n = 3(J_2-J_1)$ if $n=1$ or $n=N$ and $\alpha_n = 6J_2$ otherwise. 
Let us define Fourier transforms for $\sigma_{in}^z$, $\sigma_{in}^x$ and
$J({\bf r}_j -{\bf r}_i, m-n)$ 
in the same way as in (\ref{Fourier1}) and (\ref{Fourier2}). We now have:
\begin{equation}
{\cal H}_{ex} = \sum_{nm,{\bf q}} \left( J({\bf q}, m-n) + \alpha_n \delta_{nm}  \right) 
\left( {\sigma_{{\bf q}n}^z}^*\sigma_{{\bf q}m}^z +
(-1)^{n}{\sigma_{{\bf q}n}^x}^*(-1)^{m}\sigma_{{\bf q}m}^x \right) \ .
\end{equation}

Let us now turn to ${\cal H}_d$ and expand $\sigma_{in}^y$. Because of the layer geometry
$Q_{xz}=Q_{yz}=0$. Furthermore, since $Q({\bf q}={\bf 0})$ is diagonal,
non-diagonal terms $Q_{yx}({\bf r}_j-{\bf r}_i)\sigma_{in}^y \sigma_{jn}^x$
do not contribute at the quadratic
order, but rather yield a term linear in $\sigma_{in}^x$ which vanishes
when summed on $j$. We are thus left with diagonal terms only. The contributions of $Q_{xx}$ and $Q_{zz}$
are then:
\begin{equation}
\sum_{{\bf q}n}Q_{zz}({\bf q})|\sigma_{{\bf q}n}^z|^2 +Q_{yy}({\bf q}) |\sigma_{{\bf q}n}^x|^2
\end{equation}
and that of $Q_{yy}$, after expanding $\sigma_{in}^y$ to first order, is:
$$
-Q_{yy}({\bf q}={\bf 0})\sum_i ({\sigma_{in}^z}^2 + {\sigma_{in}^x}^2) \ = \
{\cal C} \sum_{{\bf q}n} (|{\sigma_{{\bf q}n}^z}|^2 + |{\sigma_{{\bf q}n}^x}|^2)
$$

Since only the ${\bf q}={\bf 0}$ mode contributes to $\sigma_z$, we now restrict ${\cal H}_d$
to its ${\bf q}={\bf 0}$ part. After using (\ref{Qdiag1}) and (\ref{Qdiag2}) we get
\begin{equation}
{\cal H}_d = 3{\cal C}\sum_n {\sigma_{{\bf 0}n}^z}^2
\label{dipolezpart}
\end{equation}
and 
\begin{equation}
{\cal H}_{ex} = \sum_{mn}\left(J({\bf 0},m-n)+\alpha_n\delta_{nm}\right)
\left( \sigma_{{\bf 0}n}^z\sigma_{{\bf 0}m}^z +(-1)^n\sigma_{{\bf 0}n}^x(-1)^m\sigma_{{\bf 0}m}^x
\right) . 
\end{equation}

We only need diagonalize the $z$ part of ${\cal H}$ which reads:
\begin{equation}
{\cal H}_z = \sum_{nm} \ B_{nm} \ \sigma_{{\bf 0}n}^z\sigma_{{\bf 0}m}^z
\end{equation}
with
\begin{equation}
B_{nm}= (3{\cal C} + \alpha_n)\delta_{nm} + J({\bf 0},m-n)
\end{equation}
Matrix $B$ has the same form as matrix $A$ used in section \ref{meanfield} with the difference that the matrix
elements at both ends of the principal  diagonal are different. B can be  diagonalized in the basis
of orthonormal vectors $U_k$ defined by:
\begin{equation}
(U_k)_n = \frac{1}{N_k}\sin\left((n-{1\over 2})\frac{k\pi}{N}\right)
\end{equation}
where the integer $k$ ranges from 1 to $N$. The normalization factor $N_k$ is equal to $\sqrt{N}$
if $k=N$ and $\sqrt{N/2}$ otherwise. The corresponding eigenvalues are:
\begin{equation}
b_k= 3{\cal C} + 6(J_1+J_2)\left(1+\cos(\frac{k\pi}{N})\right) \ .
\end{equation}
Expanding  $\sigma^z$ on the basis of the $\sigma_{{\bf 0}k}^z$ defined by:
$$
\sigma_{{\bf 0}k}^z = \sum_n \ (U_k)_n \ \sigma_{{\bf 0}n}^z
$$
we get: 
$$
\sigma^z = \sum_{k} \frac{\lambda_k}{N_k} \sigma_{{\bf 0}k}^z
$$
where

\begin{equation}
\lambda_k = \left\{ \begin{array}{ll}
0 & \mbox{if $k$ even} \\
\frac{\sqrt{{\cal N}}}{\sin(\frac{k\pi}{2N})} & \mbox{if $k$ odd.}
\end{array}
\right.
\end{equation}

The integration over the $\sigma_{in}^z$ in eq. (\ref{chiT}) is straightforward since the
$\sigma_{{\bf 0}k}^z$ have Gaussian weights
and we get the following result for $\chi_{out}$ at low temperature: 
\begin{equation}
\chi_{out} = \frac{(g\mu_b)^2}{2N^2}  \sum_{k=1}^N \frac{f_k}{3{\cal C} +
6(J_1+J_2)\left(1+\cos(\frac{k\pi}{N})\right)}
\label{Chifinal}
\end{equation}
where the weights $f_k$ are given by:

\begin{equation}
f_k = \left\{ \begin{array}{ll}
0 & \mbox{ if $k$ even} \\ [5pt]
\frac{2}{\sin^2(\frac{k\pi}{2N})} & \mbox{if $k$ odd and $k\neq N$} \\ [5pt]
\frac{1}{\sin^2(\frac{k\pi}{2N})} & \mbox{if $k$ odd and $k = N. $}
\end{array}
\right.
\end{equation}

The energy $\cal C$ is defined in (\ref{defineC}) and equals $ 2.759 \frac{(g\mu_b)^2}{a^3}$.
The $f_k$ satisfy the following sum rule: $\sum_k f_k =N^2$.

The $\chi_{out}$ we have found is temperature-independent; it is in fact the first term in
an expansion of $\chi_{out}$ in powers of $T$. 
Using bulk values for the couplings, we find that for a 3-layer system $\chi_{out}= 0.025 \frac{(g\mu_b)^2}{|J_1|}
$
in excellent agreement with the simulation result in Fig. \ref{Chime+si}:
$(0.026 \pm 8.10^{-4})\frac{(g\mu_b)^2}{|J_1|}$.  As the number of layers
is increased, $\chi_{out}$ slowly decreases. For a 7-layer system, expression (\ref{Chifinal}) yields:
$\chi_{out}=0.017 \frac{(g\mu_b)^2}{|J_1|} $.

\section {Discussion and conclusion}
\label{discuss}

The simulation results reported in section \ref{Montecarlo} for a 3-layer system
qualitatively reproduce the anisotropy
in susceptibilities observed in the experiments. However they do not agree quantitatively with the experimental
results. We list below
the experimental and theoretical values of $T_N$ and $\chi_{out}$ at low temperature, in dimensionless units.
\begin{eqnarray*}
\begin{array}{lll}
T_N^{exp} = 16.1 & T_N^{mc} = 9.5 \pm 0.5 & \ \ \ \ \mbox{in units of $\frac{|J_1|S^2}{k}$} \\ [5pt]
\chi_{out}^{exp} = 0.013 \ \ \ & \chi_{out}^{mc} = 0.026 \pm 8.10^{-4} & \ \ \ \ \mbox{in units of $\frac{(g\mu_b)^2}{|J_1|}$}
\end{array}
\end{eqnarray*}
Our Monte Carlo simulations have been run on finite systems: although $T_N$ and $\chi_{out}$
didn't change appreciably when we increased the size of the system to 29$\times$29 or decreased it
to 13$\times$13, we cannot rule out finite-size corrections.
For $\chi_{out}$, however, analytic and simulation results are in excellent agreement
which indicates that finite-size corrections are not significant in the low-temperature region.
The numerical values listed above suggest that the effective value of $J_1$ in the samples is 
approximately twice as big as in the bulk. However, since the simulation uses bulk values for the ratios
$J_2/J_1$ and $(g\mu_b)^2/(a^3 J_1)$ this would imply that $J_2$ and the dipolar coupling are
rescaled by the same factor. This cannot be the case, as the value of the dipolar coupling
only depends
on the in-plane lattice parameter, known from X-ray spectra. 
One possible explanation is that the exchange constants are different in the SL's relative
to bulk values. 
Because of a 2.1\% lattice mismatch between EuTe and PbTe, the SL will be strained and
the in-plane and out-of-plane lattice constants will be different from one another, and different from their bulk values
\cite{chen96}. As a result, the exchange couplings will also be somewhat different. 
Within the family of Eu chalcogenides the lattice constant increases as the size of the anion increases
from O to Te which makes it possible to study the dependence of the exchange couplings on the lattice
parameter.\cite{Zinn76} In our case the in-plane lattice constant is reduced with respect to the bulk value
while the out-of-plane constant is increased. Thus, in the SL's, the in-plane $J_1$ is likely to increase,
while the out-of-plane $J_1$ will decrease. 

Although dipolar interactions account for the flatness of $\chi_{out}$ at low temperature, we can expect
a single-ion anisotropy term of the form $\kappa\sum_{in}{S_{in}^z}^2$ to have the same effect. Let us replace
${\cal H}_d$ with such a term. The calculation of $\chi_{out}$ is similar, except that in
equation (\ref{dipolezpart}), $3\cal C$ is replaced by $\kappa$. Thus for single-ion anisotropy to have the same
effect as dipole interactions, $\kappa$ would have to be of the
order of $0.2$K, which is larger than the exchange couplings, an unlikely situation. 

In conclusion, we have presented experimental susceptibilities of EuTe/PbTe short-period
antiferromagnetic superlattices. We suggest that dipolar interactions may stabilize
long-range order in these 2D structures. Additional theoretical work along with 
more precise susceptibility data and  specific heat measurements are needed to confirm this hypothesis
and to study the critical behavior, an aspect not touched upon in the present work.



\section*{Acknowledgements}

The authors are indebted to Profs M.S. and G. Dresselhaus, A. Billoire, Th. Jolic{\oe}ur for many discussions
and suggestions. They wish to thank Profs G. Springholz and G. Bauer for making the samples
used in the present study available and are grateful to the Pittsburgh Supercomputing Center for a computing time
allocation under grant no. DMR-95-0022P.
L.B. acknowledges support from NATO under Research Grant no. 3B94FR and wishes to thank Prof. M.S. Dresselhaus
and MIT for their hospitality as well as CEA for its support. Research at MIT was supported in part by NSF grant
DMR-95-10093.

   

\newpage

\newpage

\newpage

\newpage

\newpage

\newpage

\newpage

\end{document}